\documentclass[manuscript, nonacm=true]{acmart}
\usepackage{booktabs} 
\usepackage{algorithm, algorithmic}
\usepackage{caption}
\usepackage{subcaption}
\usepackage{mathtools}

\begin{document}

\title{COVID-19 Contact Tracing: Eight Privacy Questions Explored}

\subtitle{A reply to de Montjoye et al.}

\author{Hugh Lawson-Tancred}
\orcid{}
\email{}
\affiliation{%
  \institution{Birbeck, University of London}
  \department{Institute for Data Analytics}
  \city{London}
  \postcode{WC1E 7HX}
  \country{UK}
}

\author{Henry C. W. Price}
\orcid{}
\email{}
\affiliation{%
  \institution{Imperial College}
  \department{Centre for Complexity Science and Theoretical Physics Group}
  \city{London}
  \postcode{SW7 2AZ}
  \country{UK}
}

\author{Alessandro Provetti}
\orcid{0000-0001-9542-4110}
\email{ale@dcs.bbk.ac.uk}
\affiliation{%
  \institution{Birbeck, University of London}
  \department{Department of Computer Science and Information Systems}
  \city{London}
  \postcode{WC1E 7HX}
  \country{UK}
}

\renewcommand{\shortauthors}{Lawson-Tancred et al.}

\begin{abstract}
We respond to a recent short paper by de Motjoye et el. on privacy issues with Covid-19 tracking. 
Their paper, which we discuss here, is structured around three ``toy protocols'' for the design of an app which can maximise the utility of contact tracing information while minimising the more general risk to privacy. 
On this basis, the paper proceeds to introduce eight questions against which they should be assessed. 
The questions raised and the protocols proposed effectively amount to the creation of a game with different categories of players able to make different moves. 
It is therefore possible to analyse the model in terms of optimal game design.
\end{abstract}

\maketitle

\section{Introduction}
The current literature on Covid-19 seems unanimous in recognizing that, as no treatment is yet available, the only tools that we can currently deploy to stop the epidemic are contact tracing, social distancing, and quarantine. 
In this work we discuss contact tracing from an Ethics of Computing point of view. 
In fact, work on Covid-19 contact tracing, e.g., \cite{Fer20} is mostly focussed on ``algorithmic'' instantaneous contact tracing assisted by a mobile phone application. 
The DP-3T project by \cite{DP3T} is an example of research and implementation effort that addresses this urgent need. 

Recently, a short paper by \cite{deM20} has introduced and examined three toy protocols that summarise the privacy questions surrounding the Covid-19 tracing. 
They describe the following 3 toy protocols for discussion:

\subsection*{Toy protocol 1: using location}
Each app only records its own location. 
As a result,

\begin{itemize}
	\item When a user reports as infected, they send their trajectory (location and time) to the authority.
	
	\item The authority, e.g., the NHS in Britain, shares the pseudonymous trajectories of all infected users with every user.
	
	\item Users can then check if they were in close contact with an infected individual.
\end{itemize}

\subsection*{Toy protocol 2: using Bluetooth.}
Each app broadcasts a unique identifier assigned by the authority through Bluetooth. So

\begin{itemize}
	\item When two phones are near to one another, they exchange these identifiers.

	\item When a user reports as infected, they send all the identifiers they encountered to the authority.

	\item The authority will contacts all the users whose identifier was encountered by an infected user.
\end{itemize}

\subsection*{Toy protocol 3: again using Bluetooth.}
In this version each app broadcasts a unique identifier using Bluetooth, assigned by the authority. 
This unique identifier is reset every hour.
Thanks to temporary indentifiers,

\begin{itemize}
	\item When two phones are near to one another, they exchange these identifiers.
	
	\item When a user reports as infected, they send all the identifiers that they have used (one per hour) to the authority.

	\item The authority shares the identifiers of all infected users with every user. Users can then check if they encountered one of these identifiers recently.
\end{itemize}

\subsection{The Questions}
\cite{deM20} discuss the following privacy questions, all of them related to the introduction of Covid-19 algorithmic contact tracing.
 
\begin{enumerate}
	\item  How do you limit the personal data gathered by the authority?

	\item  How do you protect the anonymity of every user?

	\item  Does your system reveal to the authority the identity of users who are at risk?

	\item  Could your system be used by users to learn who is infected or at risk, even in their social circle?
	
	\item  Does your system allow users to learn any personal information about other users?

	\item  Could external parties exploit your system to track users or infer whether they are infected?

	\item  Do you put in place additional measures to protect the personal data of infected and at risk users?
	
	\item  How can we verify that the system does what it says?
\end{enumerate}

\section{Background}

The overall purpose of the game is to maximise (pure) contact information while at the same time minimising (non-contact) personal information. 
As a typical user, Alice should know that she has been in contact with somebody (in fact Bob) without knowing that her contact is indeed Bob or indeed knowing anything else at all about Bob. 
Her ignorance of the latter two facts is as important on privacy grounds as her knowledge of the first on grounds of safety \cite{deM18}.

Contact information, however, can connect with personal information in two ways. 
The first way is that (i) personal information is needed to establish contact information \cite{Nat20a}, and the second is that (ii) personal information can relatively easily be derived from contact information \cite{Gong14}. 
The ideally-designed game will enable the exchange of contact information without such information having to be established on the basis of any more general personal information or being the possible source of a derivation of such more general personal information. 

Contact tracing should not involve any sharing with either other users or the authority of trajectory/social graph information (from which identification is possible either by the authority or by the adversary).
Given these overall parameters, it seems possible that both the protocols and the questions/answers in the paper could be advantageously altered.

Secondly and more importantly, there is considerable scope for changing the proposed protocols in order to enable optimisation of the data flow objectives.
On this basis, the first protocol needs to be amended so that the authority does not reveal the entire trajectory of an infected user to all non-infected users. 

To avoid such undesirable disclosure, the authority has to know the trajectories of all users, whether infected or not. 
This arrangement, by which all information is accumulated with the authority and the minimum possible disclosed to users, could be protocol ``1a.'' 
It seems to be a limitation of trajectory-based apps (and therefore a reason for preferring identifier-based apps) that with them it is not possible to avoid disclosure of entire trajectories to either other users or the authority (or both). 
A decision therefore has to be taken, on this point, between dispersion and aggregation of information.

This is presumably at least part of the motivation for considering identifier-based protocols; however the differences between the second and third protocols could also be clarified further. 
There are two such differences. 
The first is that protocol 2 has a fixed identifier, whereas protocol 3 has a variable identifier (to use a suboptimal term). 
This is the more conspicuous difference, and it plays a larger part in the response to the questions \cite{NCSC}. 
The second difference, however, is that protocol 2 also sends its full history of identifier encounters to the authority, whereas article 3 only sends its identifier change record. 
In the case of protocol 3, the authority is not able to figure out for itself the now at risk users. 
So it sends the variable identifiers whose status has changed to infect. 
This seems to decrease the privacy of the infected users while increasing that of the non-infected users (both at risk and risk-free), see, e.g., \cite{KumDan14} for early results on de-anonymisation. 
This difference also seems to have a material effect on the privacy vulnerability. 
For example, does it reduce or increase the knowledge of the authority about trajectories/social graphs of either group? 
If the authority is able to connect the varying identifiers, then it acquires a finer grained level information about the relevant users. 
Protocol 3 is, therefore, in effect a bet on the inability of the authority to spot the continuities in series of variable identifiers.

There seems to be a further assumption built into protocol 3. 
Any information available either to the authority or to any or all users is in principle also potentially available to the adversary. 
The existing game model thus makes the further assumption that identifier information on mobile phones is more open to hacking than trajectory information. 
However, such an assumption is not immune to challenge. 
It needs to be clarified what are the relative strengths and weaknesses of the protocols with respect to the adversary.

Thirdly, the answers to the specific questions also need to be reviewed (partly in the light of the queries about the protocols). 
We now look in turn at some of the ways in which the answers to the eight questions could be changed.

\section{A response to the questions}

\subsection*{Question 1.} 
Protocol 1 obviously discloses the whole trajectory of infected users to the authority. 
The disclosure also to non-infected users can be avoided, but only at the cost of non-infected users also revealing their entire trajectories to the authority. 
This would be the move from protocol 1 to protocol 1a. 
If there is perceived to be an inverse connection between threat status and privacy entitlement, then it would seem that this change of trajectory-based protocol would be unfair. 
A larger number of users who do not constitute a threat would see their privacy eroded in order to protect the privacy of the smaller number who have become infected. 
(It should be noted that this objection applies irrespective of whether or not any form of blame is to be attributed to the change to infected status (e.g. by disregarding social distancing etc).

In terms of the identifier-based protocols, the improvement provided by protocol 3 over protocol 2 depends on the authority not being able to reconstruct the pseudonymous social graph across the changes of identifier. 
As discussed in connection with the protocols, however, it is not obvious that it will not be possible for the authority to do that.

\subsection*{Question 2.} 
The same objection to the greater innocuousness of protocol 3 over protocol 1 and protocol 2 arises as with question 1. Presumably if the game is to rely on "special measures" to "limit the risk", then those special measures should be applied at the level of the authority not the users (where they can more easily be circumvented and less easily monitored). 
It could also be argued that re-identification by either the authority or other users automatically raises the risk of re-identification by the adversary.

\subsection*{Question 3.} 
Protocol 1 does indeed give the right answer on this question, but only at the cost of giving too much information to non-infected users. 
Again, the difference between protocol 2 and protocol 3 is not clear. 
It is a reasonable supposition that the identities of the infected group are more sensitive than those of the non-infected. 
The former are (presumably) less numerous, but they are more vulnerable to potential stigmatisation/vigilantism. 
If that assumption is right, it would form a strong objection to protocol 1. 
What this suggests is that a more nuanced distinction needs to be drawn between the type of threat posed by the authority and that posed by other users.

\subsection*{Question 4.} 
Protocol 1 obviously fails this test in its existing form, but that can be prevented by letting the authority know the trajectories of non-infected users (protocol 1a again). 
Protocol 3 seems to be definitely worse than protocol 2 on this question, however it is tweaked. 
This reinforces the suspicion that protocol 3 is not preferable to protocol 2 in any respect. 
Given the presumably greater practical difficulty of deploying protocol 3, this would seem to be a very solid grounds for rejection of protocol 3 in general.

\subsection*{Question 5.} 
This seems to be the crucial question for the overall objective of the game as outlined at the start. 
It is not obvious how the protocols can be structured to enable either users or the authority to gain only and exclusively specific contact information without either supporting it with more general personal information or creating a situation in which wider personal information can be triangulated from the contact information. 
This highlights exactly what any privacy-secure contact tracing must achieve: the Holy Grail is pure contact information, uncontaminated by any non-contact personal information.

\subsection*{Question 6.} 
This seems to raise again the question whether concentrating information with users or with the authority constitutes the greater security risk. 
Most recent major hacks have focused on concentrations of data, suggesting the more disaggregation the better. 
Hacks from large numbers of dispersed users have been less effective, so far as can be known.

\subsection*{Question 7.} 
It is not clear what additional protections could be made available. 
One possibility would be some kind of time limitation of the information or preventions on its further disclosure. 
Presumably this would in practice be a question about encryption rather than game design. 
The other obvious way to develop such protections would be through legal/regulatory constraints, but they would also clearly fall outside the scope of the intrinsic app-modelling game under consideration.

\subsection*{Question 8.} 
As in many other areas of data protection, the may be a trade-off between the transparency of the system and its privacy-protection and/or security \cite{deM13}. 
It may not be possible simultaneously to optimise all three parameters. 
On the other hand, a possible way of at least partially squaring this circle is that some form of blockchain might be deployed here.

\section{Conclusions}
The game-design approach of \cite{deM20} seems to offer, perhaps with some modification, a good basis for a design intended to optimise the protection of the relevant moral assets. 
Non-infected users have an interest in learning about vulnerability-increasing contacts, but have no right to any other information about infected users. 
Infected users have a duty to maximise knowledge about their contacts, but the right not to have any further information about them disclosed. 
(As we have seen, the right of the infected might outweigh that of the non-infected on the grounds of the risk of vigilantism, whereas it might also be thought to be outweighed because of the possible culpability of infection given the level of public knowledge \cite{Bro20}. 
This is clearly a value, not a pure design, issue.) 
The authority has the right (and possibly duty) to be as informed as possible about the pattern of spread of the epidemic, but it should be prevented from acquiring (and indeed retaining) any more than the essential information about the users under its jurisdiction. 
The adversary has no rights in this context and is simply a threat to be minimised.

The most significant improvement to the approach proposed, in our opinion, would be to replace protocol 1 with our protocol 1a. 
We are agnostic as to the general preference for trajectory/identifier approaches, but we suspect that with the latter protocol 3 on balance creates a greater privacy risk than protocol 2.

\begin{acks}
This article is based upon work from COST Action DigForAsp CA17124, supported by COST (European Cooperation in Science and Technology): \url{www.cost.eu}.
\end{acks}

\bibliographystyle{ACM-Reference-Format}
\bibliography{contact-tracing-arXiv}


\begin{thebibliography}{10}


\ifx \showCODEN    \undefined \def \showCODEN     #1{\unskip}     \fi
\ifx \showDOI      \undefined \def \showDOI       #1{#1}\fi
\ifx \showISBNx    \undefined \def \showISBNx     #1{\unskip}     \fi
\ifx \showISBNxiii \undefined \def \showISBNxiii  #1{\unskip}     \fi
\ifx \showISSN     \undefined \def \showISSN      #1{\unskip}     \fi
\ifx \showLCCN     \undefined \def \showLCCN      #1{\unskip}     \fi
\ifx \shownote     \undefined \def \shownote      #1{#1}          \fi
\ifx \showarticletitle \undefined \def \showarticletitle #1{#1}   \fi
\ifx \showURL      \undefined \def \showURL       {\relax}        \fi
\providecommand\bibfield[2]{#2}
\providecommand\bibinfo[2]{#2}
\providecommand\natexlab[1]{#1}
\providecommand\showeprint[2][]{arXiv:#2}

\bibitem[\protect\citeauthoryear{{de Montjoye}, {Hidalgo}, {Verleysen}, and
  {Blondel}}{{de Montjoye} et~al\mbox{.}}{2013}]%
        {deM13}
\bibfield{author}{\bibinfo{person}{Yves-Alexandre {de Montjoye}},
  \bibinfo{person}{C{\'e}sar~A. {Hidalgo}}, \bibinfo{person}{Michel
  {Verleysen}}, {and} \bibinfo{person}{Vincent~D. {Blondel}}.}
  \bibinfo{year}{2013}\natexlab{}.
\newblock \showarticletitle{{Unique in the Crowd: The privacy bounds of human
  mobility}}.
\newblock \bibinfo{journal}{\emph{Scientific Reports}}  \bibinfo{volume}{3},
  Article \bibinfo{articleno}{1376} (\bibinfo{date}{March}
  \bibinfo{year}{2013}), \bibinfo{numpages}{1376}~pages.
\newblock
\urldef\tempurl%
\url{https://doi.org/10.1038/srep01376}
\showDOI{\tempurl}


\bibitem[\protect\citeauthoryear{de~Montjoye, Houssiau, Gadotti, and
  Guepin}{de~Montjoye et~al\mbox{.}}{2020}]%
        {deM20}
\bibfield{author}{\bibinfo{person}{Yves-Alexandre de Montjoye},
  \bibinfo{person}{Florimond Houssiau}, \bibinfo{person}{Andrea Gadotti}, {and}
  \bibinfo{person}{Florent Guepin}.} \bibinfo{year}{2020}\natexlab{}.
\newblock \bibinfo{booktitle}{\emph{Evaluating COVID-19 contact tracing apps?
  Here are 8 privacy questions we think you should ask}}.
\newblock \bibinfo{type}{{T}echnical {R}eport}. \bibinfo{institution}{Imperial
  College, Computational Privacy Group}.
\newblock
\urldef\tempurl%
\url{https://cpg.doc.ic.ac.uk/blog/evaluating-contact-tracing-apps-here-are-8-privacy-questions-we-think-you-should-ask/}
\showURL{%
\tempurl}


\bibitem[\protect\citeauthoryear{Ferretti, Wymant, Kendall, Zhao, Nurtay,
  Abeler-D{\"o}rner, Parker, Bonsall, and Fraser}{Ferretti
  et~al\mbox{.}}{2020}]%
        {Fer20}
\bibfield{author}{\bibinfo{person}{Luca Ferretti}, \bibinfo{person}{Chris
  Wymant}, \bibinfo{person}{Michelle Kendall}, \bibinfo{person}{Lele Zhao},
  \bibinfo{person}{Anel Nurtay}, \bibinfo{person}{Lucie Abeler-D{\"o}rner},
  \bibinfo{person}{Michael Parker}, \bibinfo{person}{David Bonsall}, {and}
  \bibinfo{person}{Christophe Fraser}.} \bibinfo{year}{2020}\natexlab{}.
\newblock \showarticletitle{Quantifying SARS-CoV-2 transmission suggests
  epidemic control with digital contact tracing}.
\newblock \bibinfo{journal}{\emph{Science}} \bibinfo{volume}{368},
  \bibinfo{number}{6491} (\bibinfo{year}{2020}).
\newblock
\showISSN{0036-8075}
\urldef\tempurl%
\url{https://doi.org/10.1126/science.abb6936}
\showDOI{\tempurl}
\showeprint{https://science.sciencemag.org/content/368/6491/eabb6936.full.pdf}


\bibitem[\protect\citeauthoryear{Gong, Morikawa, Yamamoto, and Sato}{Gong
  et~al\mbox{.}}{2014}]%
        {Gong14}
\bibfield{author}{\bibinfo{person}{Lei Gong}, \bibinfo{person}{Takayuki
  Morikawa}, \bibinfo{person}{Toshiyuki Yamamoto}, {and}
  \bibinfo{person}{Hitomi Sato}.} \bibinfo{year}{2014}\natexlab{}.
\newblock \showarticletitle{Deriving Personal Trip Data from GPS Data: A
  Literature Review on the Existing Methodologies}.
\newblock \bibinfo{journal}{\emph{Procedia - Social and Behavioral Sciences}}
  \bibinfo{volume}{138} (\bibinfo{year}{2014}), \bibinfo{pages}{557 -- 565}.
\newblock
\showISSN{1877-0428}
\urldef\tempurl%
\url{https://doi.org/10.1016/j.sbspro.2014.07.239}
\showDOI{\tempurl}
\newblock
\shownote{The 9th International Conference on Traffic and Transportation
  Studies (ICTTS 2014).}


\bibitem[\protect\citeauthoryear{Nat}{Nat}{2020}]%
        {Nat20a}
\bibfield{author}{\bibinfo{person}{Nat}.} \bibinfo{year}{2020}\natexlab{}.
\newblock \showarticletitle{Show evidence that apps for COVID-19
  contact-tracing are secure and}.
\newblock \bibinfo{journal}{\emph{Nature}}  \bibinfo{volume}{580}
  (\bibinfo{date}{apr} \bibinfo{year}{2020}).
\newblock
\showISSN{1476-4687}
\urldef\tempurl%
\url{https://doi.org/10.1038/d41586-020-01264-1}
\showDOI{\tempurl}


\bibitem[\protect\citeauthoryear{{Radaelli}, {Sapiezynski}, {Houssiau},
  {Shmueli}, and {de Montjoye}}{{Radaelli} et~al\mbox{.}}{2018}]%
        {deM18}
\bibfield{author}{\bibinfo{person}{Laura {Radaelli}}, \bibinfo{person}{Piotr
  {Sapiezynski}}, \bibinfo{person}{Florimond {Houssiau}}, \bibinfo{person}{Erez
  {Shmueli}}, {and} \bibinfo{person}{Yves-Alexandre {de Montjoye}}.}
  \bibinfo{year}{2018}\natexlab{}.
\newblock \showarticletitle{{Quantifying Surveillance in the Networked Age:
  Node-based Intrusions and Group Privacy}}.
\newblock \bibinfo{journal}{\emph{arXiv e-prints}}, Article
  \bibinfo{articleno}{arXiv:1803.09007} (\bibinfo{date}{March}
  \bibinfo{year}{2018}), \bibinfo{numpages}{arXiv:1803.09007}~pages.
\newblock
\showeprint[arxiv]{cs.CY/1803.09007}


\bibitem[\protect\citeauthoryear{Rosand, Koser, and Schumicky-Logan}{Rosand
  et~al\mbox{.}}{2020}]%
        {Bro20}
\bibfield{author}{\bibinfo{person}{Eric Rosand}, \bibinfo{person}{Khalid
  Koser}, {and} \bibinfo{person}{Lilla Schumicky-Logan}.}
  \bibinfo{year}{2020}\natexlab{}.
\newblock \bibinfo{booktitle}{\emph{Preventing violent extremism during and
  after the COVID-19 pandemic}}.
\newblock
\urldef\tempurl%
\url{https://www.brookings.edu/blog/order-from-chaos/2020/04/28/preventing-violent-extremism-during-and-after-the-covid-19-pandemic/}
\showURL{%
\tempurl}


\bibitem[\protect\citeauthoryear{Sharad and Danezis}{Sharad and
  Danezis}{2014}]%
        {KumDan14}
\bibfield{author}{\bibinfo{person}{Kumar Sharad} {and} \bibinfo{person}{George
  Danezis}.} \bibinfo{year}{2014}\natexlab{}.
\newblock \showarticletitle{An Automated Social Graph De-Anonymization
  Technique}. In \bibinfo{booktitle}{\emph{Proceedings of the 13th Workshop on
  Privacy in the Electronic Society}} \emph{(\bibinfo{series}{WPES 14})}.
  \bibinfo{publisher}{Association for Computing Machinery},
  \bibinfo{address}{New York, NY, USA}, \bibinfo{pages}{47--58}.
\newblock
\showISBNx{9781450331487}
\urldef\tempurl%
\url{https://doi.org/10.1145/2665943.2665960}
\showDOI{\tempurl}


\bibitem[\protect\citeauthoryear{{The National Cyber Security Centre}}{{The
  National Cyber Security Centre}}{2020}]%
        {NCSC}
\bibfield{author}{\bibinfo{person}{{The National Cyber Security Centre}}.}
  \bibinfo{year}{2020}\natexlab{}.
\newblock \bibinfo{booktitle}{\emph{The security behind the NHS contact tracing
  app}}.
\newblock
\urldef\tempurl%
\url{https://www.ncsc.gov.uk/blog-post/security-behind-nhs-contact-tracing-app}
\showURL{%
\tempurl}


\bibitem[\protect\citeauthoryear{Troncoso, Payer, Hubaux, Salathé, Larus,
  Bugnion, Lueks, Stadler, Pyrgelis, Antonioli, Barman, Chatel, Paterson,
  Capkun, Basin, Beutel, Jackson, Preneel, Smart, Singelee, Abidin, Gürses,
  Veale, Cremers, Backes, Binns, Cattuto, Persiano, Fiore, Barbosa, and
  Boneh}{Troncoso et~al\mbox{.}}{2020}]%
        {DP3T}
\bibfield{author}{\bibinfo{person}{Carmela Troncoso}, \bibinfo{person}{Mathias
  Payer}, \bibinfo{person}{Jean-Pierre Hubaux}, \bibinfo{person}{Marcel
  Salathé}, \bibinfo{person}{James Larus}, \bibinfo{person}{Edouard Bugnion},
  \bibinfo{person}{Wouter Lueks}, \bibinfo{person}{Theresa Stadler},
  \bibinfo{person}{Apostolos Pyrgelis}, \bibinfo{person}{Daniele Antonioli},
  \bibinfo{person}{Ludovic Barman}, \bibinfo{person}{Sylvain Chatel},
  \bibinfo{person}{Kenneth Paterson}, \bibinfo{person}{Srdjan Capkun},
  \bibinfo{person}{David Basin}, \bibinfo{person}{Jan Beutel},
  \bibinfo{person}{Dennis Jackson}, \bibinfo{person}{Bart Preneel},
  \bibinfo{person}{Nigel Smart}, \bibinfo{person}{Dave Singelee},
  \bibinfo{person}{Aysajan Abidin}, \bibinfo{person}{Seda Gürses},
  \bibinfo{person}{Michael Veale}, \bibinfo{person}{Cas Cremers},
  \bibinfo{person}{Michael Backes}, \bibinfo{person}{Reuben Binns},
  \bibinfo{person}{Ciro Cattuto}, \bibinfo{person}{Giuseppe Persiano},
  \bibinfo{person}{Dario Fiore}, \bibinfo{person}{Manuel Barbosa}, {and}
  \bibinfo{person}{Dan Boneh}.} \bibinfo{year}{2020}\natexlab{}.
\newblock \bibinfo{booktitle}{\emph{DP-3T: Decentralized Privacy-Preserving
  Proximity Tracing}}.
\newblock
\urldef\tempurl%
\url{https://github.com/DP-3T/documents}
\showURL{%
\tempurl}


\end{thebibliography}
\end{document}